# Innovation as Evolution

*Case of Study: Phylomemetic of Cellphone Designs*

Deni Khanafiah[1], Hokky Situngkir[2]

[1]Research assistant in Bandung Fe Institute. mail: dk@students.bandungfe.net
[2]Department Computational Bandung Fe Institute. mail: hokky@elka.ee.itb.ac.id.
Personal Web: http://www.bandungfe.net/hs

**Abstract**

Cellular phone is one of the most developing technological artifacts today. The evolution occurs through random innovation. Our effort is trying to view the evolution of this artifact from memetics. By constructing a phylomemetic tree based on cellular phone memes to infer or estimate the evolutionary history and relationship among cellular phone. We adopt several methods, which are commonly used in constructing phylogenetic tree, they are UPGMA algorithm and Parsimony Maximum algorithm to construct cellphone phylomemetic tree. Therefore we compare with the innovation tree, which is based on serial number and their appearance time. From phylomemetic tree, we then analyze the process of a cellular phone innovation through looking out on the cellular phone type lies in the same cluster. The comparison of the simulation tree result shows a generally different branching pattern, giving a presumption that innovation in cellular phone is not really relating with their serial number, but occurs merely because of random mutation of allomeme design and competes with its technological development.

## 1. The process of Innovation and Evolution of Cellular Telephone

Innovation is one of many processes that may emerge new artifacts. As a (complex) system, innovation of artifact means the change of state from the component of the system, in so forth emerging a system, which its characters or behaviors are different from the previous time (Frenken, 2001a). A process of innovation can be regarded as a relatively random process (Mokyr, 1997; Frenken, 2001a; Kauffman, 1995), in which means the resulting technology cannot always be known precisely to fit with the environment or not.

Evolution process from technological artifacts can be viewed as the phenomenon of the emerging of the new type of artifacts by mean of innovation, which in turns will replace the old ones. Up to these days, study or analysis to understand the process and the evolution of innovation is still an interesting field, especially to obtain deeper understanding of the principles behind its evolution (Kaplan et al, 2003).

One of the interesting technological artifacts to study is cellular telephone (*cellphone*). It is a technological artifact that is still evolving up to this very second. The evolution process has been emerging abundant and the more complex variants of cellphones.

Cellphone is a very fast-growing telecommunication tool. At least we can observe this from the drastically (exponential) growing numbers of the GSM type cellphone users from years 1991-2001 (Figure 1). Cellphone, as a wireless transmission technology is the root for various innovations, which enable the users to communicate in mobile way spatially (Geser, 2004). In other words, it has been growing to become a technological artifacts which have a particular functions and affects the human culture in a certain fashion (Humphreys, 2003).

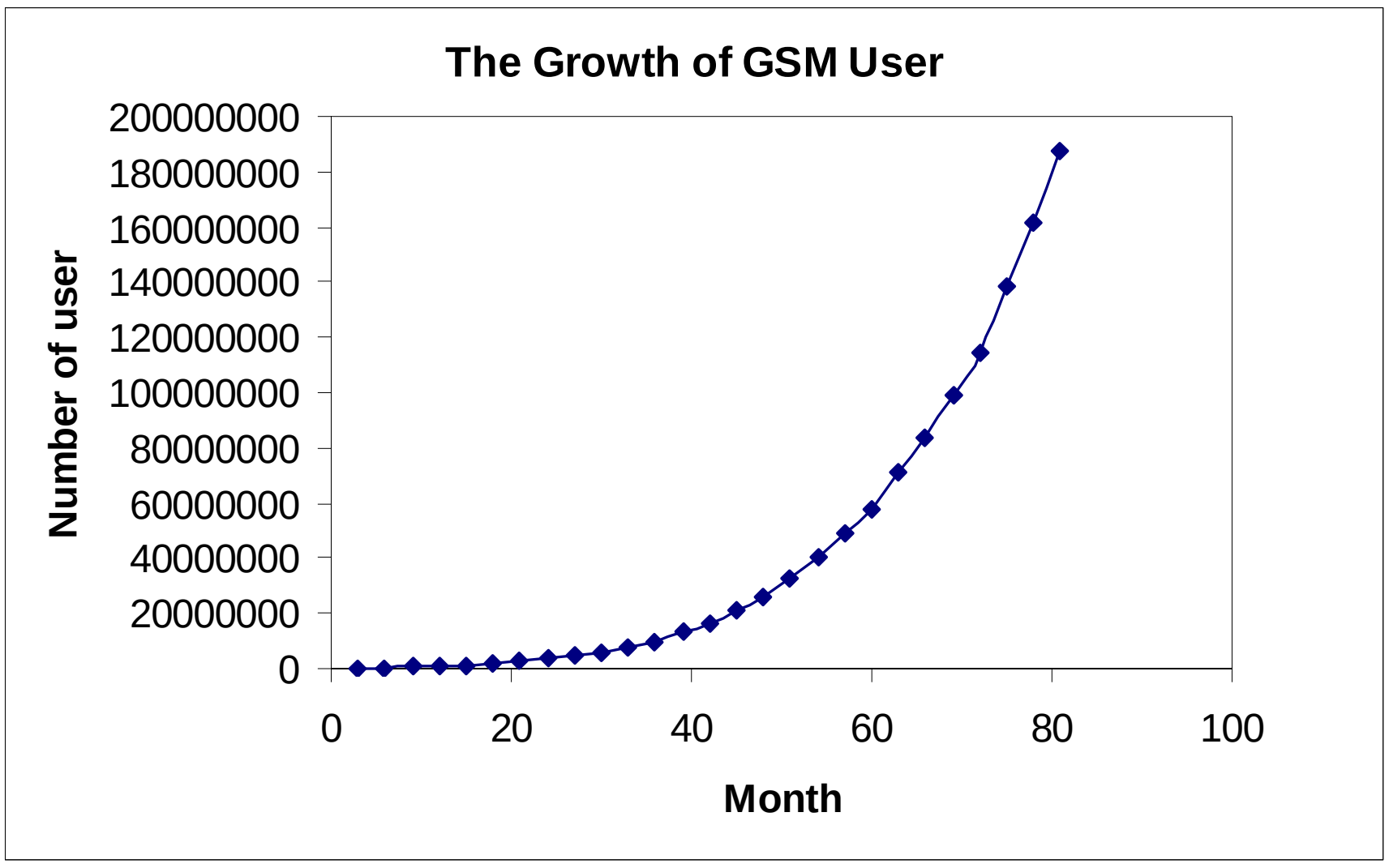

**Figure 1**
The growing number of GSM cellphone users vs Time from September 1992 to June 1999.
(data obtained from http://www.cellular.co.za/stats/statistics_global_by_standard.htm)

Cellphone is a technological artifact that changes from time to time. The change may happen over its design and technology. Evolution of cellphone design or commonly known as cellphone trend or *fashion*[1], in other words is how a new variant (new design of the different cellphones) emerge over time (Figure 2), e.g. the *cover* or the case of the *handset*, the form and placement of the antenna for instances, etc.

[1] On how cellphone development seen as *fashion* development, see review of Eldar Murtazin in website mobile-review.com (URL:http://www.mobile-review.com/review-en.shtml)

Beside their designs, cellphones, as well as other artifacts, we can see them as an assembled of components of technology, which are collectively emerging particular function (Frenken, 2001b). From technology point of view, cellphone evolution could mean the the cumulative change of its function as the result of its components change. The function of cellphone is due to their capability to resulting certain activities or functions. Be it, as telecommunication mean to gaming, camera, radio FM, Internet browsing, sending messages, etc. Shortly, the evolution of technology of cellphones is as in Figure 3.

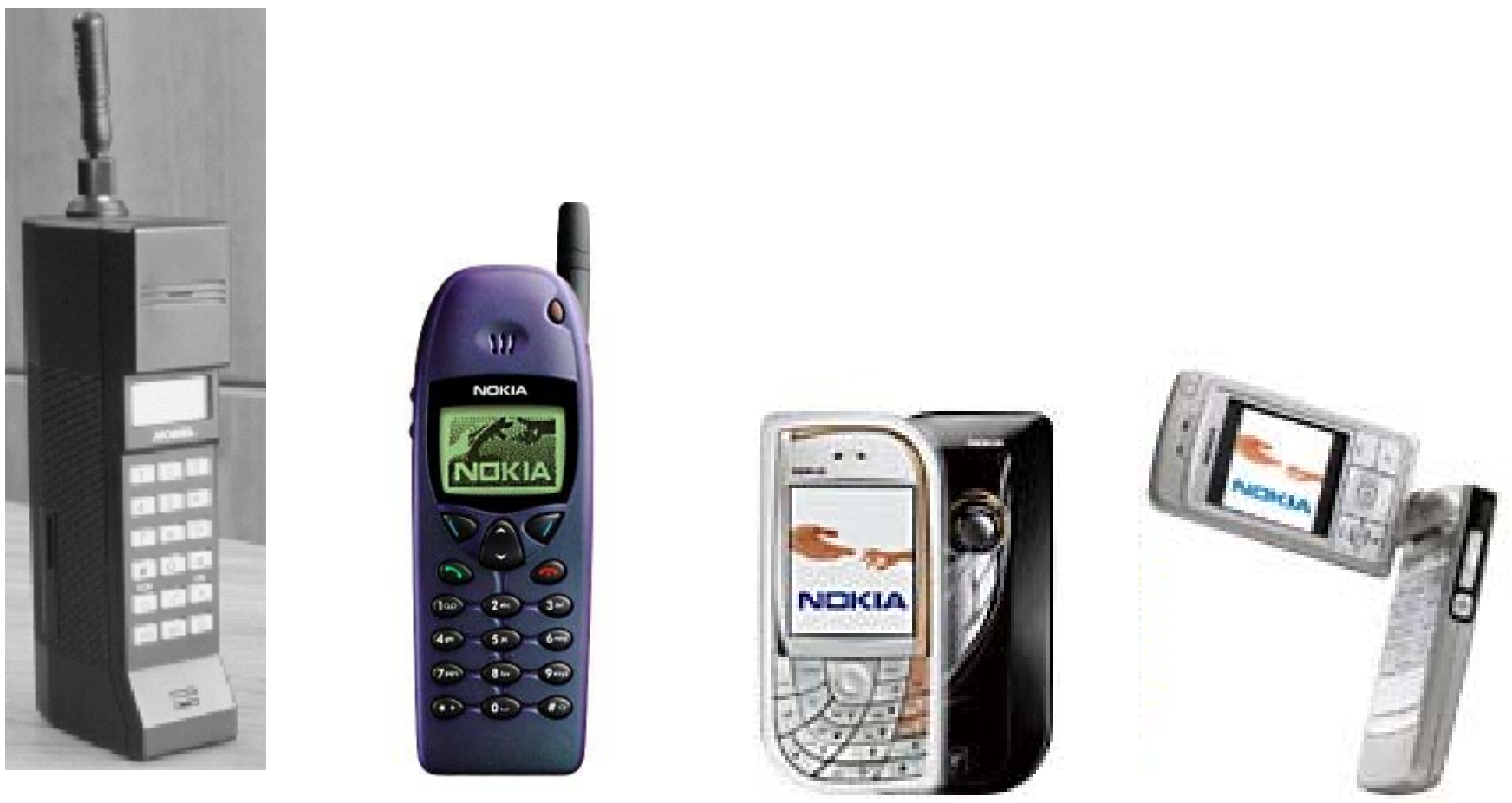

**Figure 2**
**Examples of design evolution of Nokia cellphones**. Nokia Mobira *Cityman*™ (leftest) is an 80's Nokia cellphone. Nokia 6110™ (second from left) is Nokia of the late 90's. Nokia 7610™ and Nokia 6260™ (the third and fourth from left) are few of cellphones today.

In this paper, we will see the evolution process of cellphones design and technology, using memetic perspective. Memetic is a concept that brings out Darwinian evolution paradigm – firstly developed in Biology – to understand the evolution phenomena in social system. Memetics becomes an alternative analytical tool to understand socio-culture phenomena (Situngkir, 2004), including innovation of cellphone as one of technological artifacts.

One of analyses we do in understanding cellphone evolution process with memetics is analysis of cellphone evolutionary relationship (ancestral relationship of cellphones). In this analysis, we adopt the phylogenetic concept – a concept used in Biology to infere the history and the evolutionary relationship of organisms based on their relativity characters. Using the same paradigm, we infer or estimate the evolutionary relationship of a cellphone based on cellphone memes as their comparison characters. The evolutionary relationship among different types of cellphones is represented as a branching, a tree-like diagram defined as phylomemetic tree.

## 2. Cellphone Innovation as a Memetic Evolution

As explained in Sartika (2004), memetics has many definitions since the term and the concept coined for the first time by Dawkins (1976), i.e. meme as a replicator unit (Blackmore, 1998), unit of information transmission which becomes subject of selection process (Wilkins, 1998), item of memory or a number of certain information that is stored neurally (Lynch, 1988), etc. In memetics, we consider social system or culture as a system compounded by many units of cultural evolution or the smallest selection unit be called meme (Wilkins, 1998).

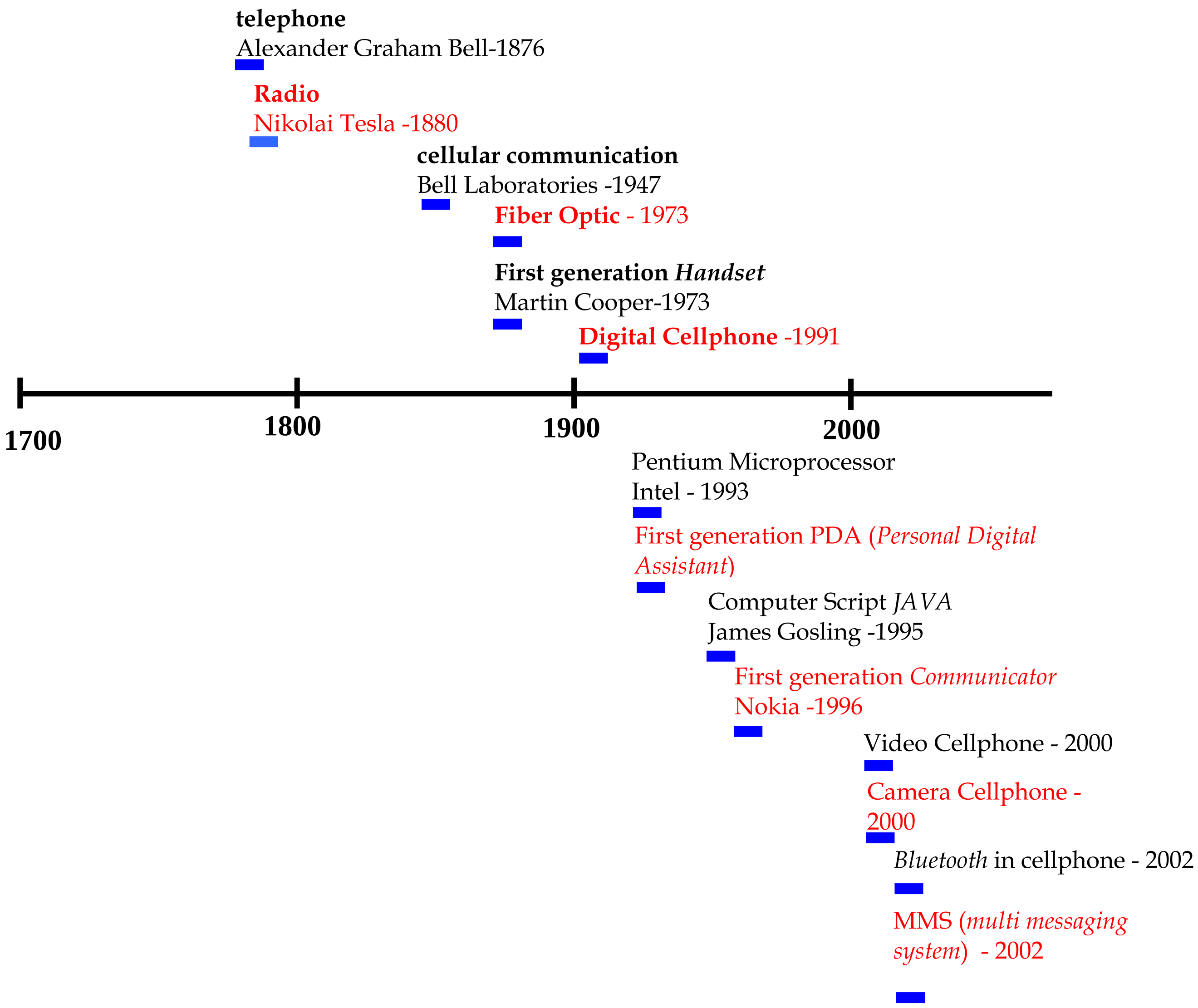

**Figure 3**
*Timeline* technological innovation that affect the cellphone evolution

Cellphones, as well as other artifacts, is one of cultural objects. It is a system which consist cultural information units as the heredity units which pass from generation to generations in their evolution (read: innovation). We can determine the character of this technological artifact by those information units that express in certain way (Stankiewicz, 2000). In other words, cellphone is a phenotypic meme or we call it femotype that emerged from its genotype, that is memetype. So we can say that the process of cellphones innovation is a memetic evolution.

Innovation in memetis perspective can be regarded as a random mutation from meme codes of artifacts. Mutation or the change of meme from one code to different code in cellphone will cause variation in cellphone as the source or material for the selection process.

It is a difficult effort to identify memes that constitute a cellphone. Since meme here is an abstract thing. However, as Situngkir cited in (2004), we can view meme from the evolutionary cultural object as the smallest unit of information we can identify and use to explain the evolution process. Cellphone memes as the smallest unit of information expresses certain characters or traits in cellphones, be it the design, the technology, the function, and other traits. We can identify cellphone meme through observing the traits or the property that the cellphones have and then determine the smallest unit of information that can be used to explain its evolutionary process.

## 3. Memetic model of cellphones

In modeling cellphone evolution in memetic perspective, we use a model that Situngkir developed in Situngkir (2004) and Situngkir *et al* (2004). In the model, we describe cellphone as a system constituted of meme or unit of information as the smallest unit of replication. The memes will compound memeplexes, where in memetic process expressed as femetype i.e. the trait of design of a cellphone. By denoted the set of memetype as $M$ and femetype as $C$, and function $\rho$ as the function that correlates $M$ with $C$. We formulate the relation as:

$$\rho : M \to C \qquad (1)$$

Generally, memetype is a memeplex constituted of a number of certain memes. Say it a memeplex constituted of $N$ memes, where each meme will have one alternative meme called alomeme ($A$), thus we can denote memeplex as:

$$M = A_1 A_2 ... A_N \qquad (2)$$

with A assumed as  set of all allomeme:

$$A = \bigcup_{i=1}^{n} A_i \qquad (3)$$

As Heylighen (1993) proposed and applied computationally in Situngkir (2004) and Situngkir *et al* (2004), allomeme in cellphone can be stated as a "yes" or "no" over proposition of certain character of cellphone that constitute in the statement "IF...THEN...". We then represent allomeme of each cellphone as binary number *(1,0)* that represents the presence of certain meme in the cellphone.

Mutation process in a cellphone memeplex will result a generation that has different set of memeplex, which emerges as an artifact with different trait. In formal, this process is written as follow:

$$I : MxC_M \to M^* \qquad (4)$$

with $I$ as mutation process in memeplex that will map its ancestor memeplex $(M)$ to the next generation memeplex $(M^*)$ with parameter of mutation control $C_M$.

In constituting the above cellphone, we use information of design features that represent a character of cellphone. We can use information of cellphone features as the basic information, which differs one type of cellphone to another.

The possible memes are:

1. IF it is infrared cellphone THEN there will be infrared system
2. IF it has SMS (short message service) function THEN there will be sms technology
3. And so forth.

There will be a lot of memes that represent the character of a cellphone depends on the data that we have. In this paper, we use cellphone Nokia™ for a case study. We base the arrangement of memeplex in cellphone Nokia™ on the feature information of each type in the official website of the cellphone[2]. From the data we obtained, we pick 66 types of cellphone. We modeled the memeplex for each type based on feature information of each variant. Thus, we identified 84 characteristic, be it design, or technological function that can possibly be the constitute memes of the memeplex set of Nokia cellphone (lsee appendix 1).

## 4. Construction of Cellphone Phylomemetic Tree – Inference of Evolutionary Relationship of Technological Artifacts

Evolution is a gradual process, where simple species evolves to become more complex species through accumulation of character change inherited from generation to generation. A descendent will have several different properties from its ancestor because it is changing in its evolution (Estabrook, 1987). Systematics is a branch in Biology that studies genealogical relationship among organisms and also tries to describe the pattern of evolutionary events, which causes certain distribution and diversity in living things. Systematic analysis is conducted through constructing the history of evolution and the evolutionary relationship between descendents to their ancestors based on the similarities of characters as the basic of comparison (Lipscomb, 1998). This kind of analysis is well known as phylogenetic analysis, or sometimes called Cladistics, which means a 'clade' or set of descendents from one common ancestor. Phylogenetic analysis often represented as a branching system, a tree-like diagram known as phylogenetic tree (Brinkman, 2001).

Evolution of technological artifacts, as in cellphone, is of course different with the evolution in biology system. However, the same paradigm may become an alternative analysis to understand the innovation process in cellphones.

### 4.1 Phylogenetic Tree

In living systems, evolution process involves genetic mutation and recombination process in a species so that it is resulting a new different species. Evolutionary history of an organism can be identified from the change of its characteristic. The similar characteristic is the base of analyzing the relationship of one species from other species.

In this case, tree diagram is a logical way to show the evolutionary relationship among organisms (Schmidt, 2003). Phylogenetic is a model that represents the approximate ancestral relationship of organisms, the sequence molecules, or both (Brinkman, *et al*, 2001). Phylogenetic for the organism we study, is the diagram that represents continuity of genealogy of an organism from time to time, where a point of branch showing a divergence while number of lines on certain time represents number of *taxa* at the interval time.

---

[2] alamat website: http://www.nokia.com. Pemilihan didasarkan semata-mata karena tersedianya data yang relatif lebih banyak mengenai tipe-tipe telepon selularnya serta fitur-fitur yang dimilikinya. Tentu saja analisis dapat kita lakukan pada artefak yang lainnya, yang sangat bergantung pada ketersediaan data yang cukup detail dan memadai mengenai artefak tersebut.

The arrangement of phylogenetic tree has several objections, they are: to construct precisely genealogical relationship among organisms and to estimate the event where divergence occurs from one ancestor to their descendents (Li, *et al,* 1999).

In a graph form, we formally define tree as an acyclic graph $T = (V, E)$, where $V$ is set of nodes and $E$ is set of edges set that connect two nodes of the graph. Phylogenetic as binary tree is a graph which has all node of degree one or three (Waterman, 1995). The degree of a node is defined as number of lines connected to the node.

In evolutionary tree (Figure 4), each node will represent a species or taxa. The lines or we call them braches describe the evolutionary development of the nodes. The nodes in the tree is divided into 3 types for tree with roots, they are root node, internal node, and external node. Root node $\in V$ is described as the node where the lines are in outward direction. External node or *leaf node* is the node representing species or taxa. This node has a degree of node = 1. Internal node is the node that represents common ancestor of each descendent. Internal node has a degree >1. In binary tree, internal node will have a degree = 3.

## 4.2 Phylomemetic Tree

As we have explained in previous sections that inference, or estimation of evolutionary history of an organism, is one of the main goal in phylogenetic analysis. The same paradigm in that phylogenetic study has become one of the most interesting alternatives to apply in analyzing cellphone innovation when we see it as the evolving entity, in this case through memetics. The characteristics of a type of cellphone are determined the constituting memes. Innovation process as a random mutation from its memeplex (equation 4), will result a new kind of cellphone that has several different "traits" with its ancestors. We can use cellphones memes as the basic comparison in understanding evolutionary relationship among cellphones. Cellphones with close relationship will have similar characteristic with the distance cellphone.

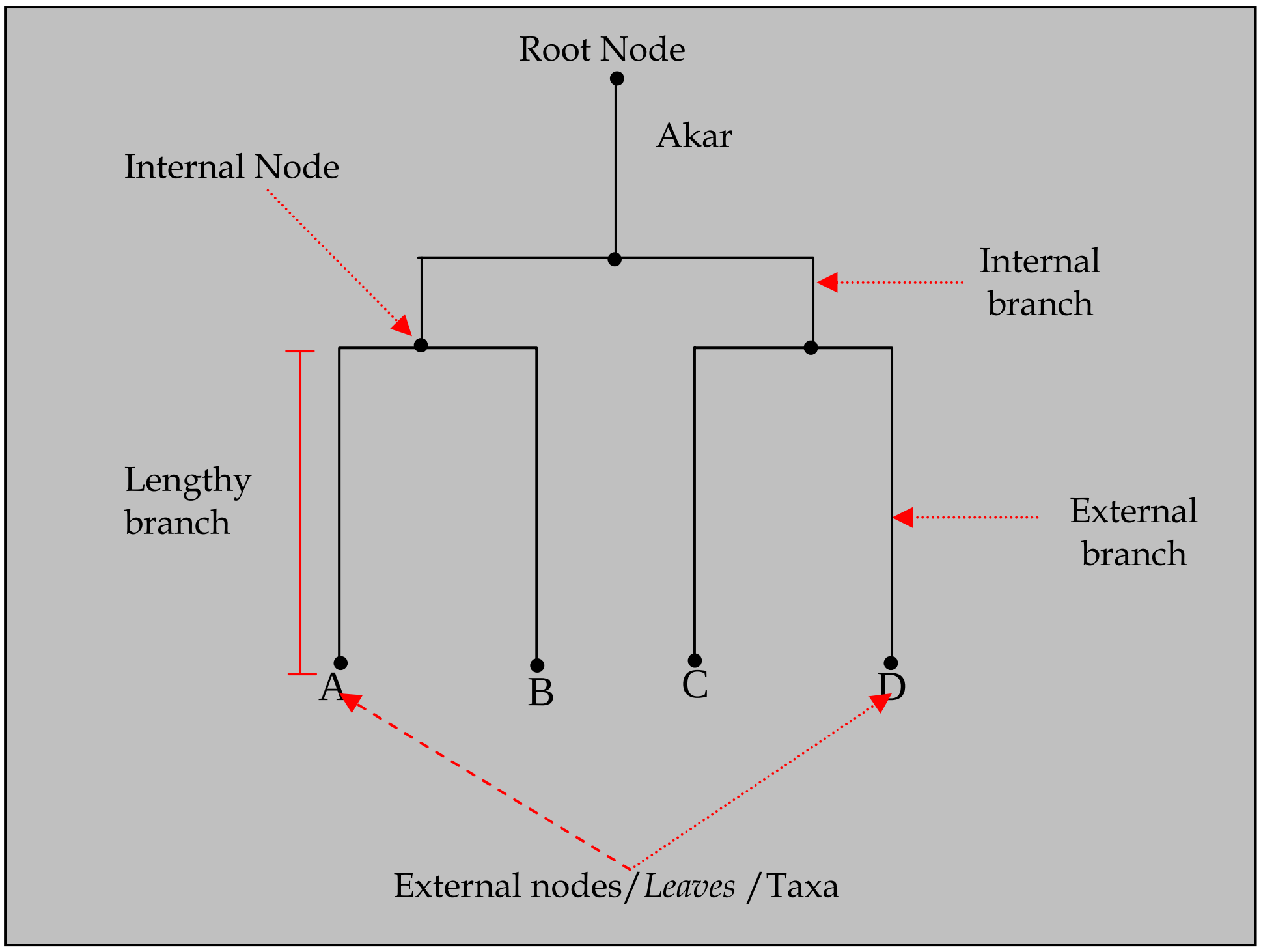

**Figure 4**
Rooted phylogenetic tree

Using the same representation of phylogenetic, in the form of tree diagram, we can try to describe the approximate evolution history and the evolutionary relationship among cellphones. We define the tree as a phylomemetic tree when we use the binary sequence of memes that constitutes it as the comparison characters. Phylomemetic tree is a tree-like diagram that describes evolution history and the relationship of the observing femetypes or memeplexes.

The phylomemetic tree for *N* types of cellphones with memeplex $M_i$, where $i = 1,...,N$, is an acyclic graph $T(N) = (V, E, M)$ consists set of nodes $V$ and set of edges $E$ that connect one node with memeplex $M_i$ with *i=(1...N)*, to another ones. Each external node at phylomemetic tree represents artifact that we infer its evolutionary relationship.

### 4.3 Construction of Nokia Phylomemetic Tree

Constructing phylomemetic tree is one of the ways that we can use to infer the evolutionary relationship between cellphone.. One phylomemetic tree should have been precise enough to describe evolution history of cellphones based on the input data in form of sequence of memeplexes for each cellphone. That is why we need the right method of constructing phylomemetic tree.

There have been many methods used in constructing phylogenetic tree and we can use it as an alternative method to construct cellphone phylomemetic tree. In this paper, we will see several general methods in constructing phylogenetic tree and then we will see how the same method can be used in constructing phylomemetic tree. Phylogenetic algorithm generally consists of two types, i.e.: (1) Algorithm construction using distance, including UPGMA and *Neighbor Joining* and (2) Algorithm construction based on character of: The Shortest Tree (*parsimony maximum*) and maximum likelihood (Brinkman, 2001).

We will use the method that is similar with the method to construct the above phylogenetic tree, in order to construct our cellphone phylomemetic tree. In this paper we choose 2 kinds of general method in constructing phylogenetic that represents the method based on distance, and based on character, that is UPGMA and Parsimony respectively.

#### 4.3.1 Unweighted Pair Group Method with Arithmetic Mean (UPGMA)

UPGMA or Unweighted Pair Group Method with Aritmatic Mean is a relatively simple and common method in constructing phylogenetic tree. It was developed for the first time to describe taxonomy phenogram .i.e. a tree that describes phenotypic similarity among species, which then further developed as method for construction of phylogenetic tree under assumption that each species evolves independently with the same rate (Opperdoes, 1997).

UPGMA is a tree-constructing algorithm using clustering analysis technique, which based on character similarity among units to be then represented as distance. When we choose genetic sequence, macromolecules or other sequences as the basic comparison of characters, then the distance between sequence will represent the cost mutation from one sequence to other sequences (Waterman, 1995. pp. 192).

UPGMA tries to construct a tree where species will correlate with other species that has the bigger character similarities, or has the minimum mutation distance. In UPGMA, we construct tree based on distance, where species with near distance is put in the same cluster. The distance between node of cluster *c* – that contains *i* and *j* with other unit *k*, is calculated using average system in arithmetic as follow:

$$d(c,k) = \frac{d(i,k) + d(j,k)}{2}, \text{ with } c = \{i, j\} \quad (5)$$

In short, the construction algorithm using UPGMA (Opperdoes, 1997) is as follow:

1. Choose the minimum distance between two taxa $i$ and $j$ ($d(i, j)$)
2. Joint *i* and *j* as one new node $\{i, j\} = c$
3. Calculate distance from *c* to other node *k* using equation (5)
4. Erase column and line that contains *i* and *j* then substitute with *c* ={*i,j*}
5. Repeat from step 1 until last one row and column

UPGMA will result the tree with topology or certain pattern of branching, which is hoped to have the minimum distance between nodes. This is a quite fast method to approximate the topology of phylogenetic tree. However, it has a weakness since it does not include time evolution into account, so the time when the species diverge into new species remain unknown.

In constructing cellphone phylomemetic tree using this method, we use cellphone memeplex that we have modeled previously as a comparison. To measure the distance between two memeplexes, we use *divergence distance* between two sequences (Wagner, 1984), this is often called *Manhattan* distance (Brooks, 1984). It is defined as number of different characters between two sequences divided with total number of characters.

By regarding mutation value or cost of each meme representing the change of state for each meme from *1* to *0* or in reverse equals one, so the distance between cellphones $i$ and $j$ with memeplex $M_i = a_1 a_2 ... a_k$ and $M_j = b_1 b_2 ... b_n$ which is sequence with option *(1,0)* along $k$ will value:

$$d(i, j) = \frac{jumlah\ a_i \neq b_i}{k}, \text{ with } i = 1,...,k \quad (6)$$

The resulting distance we use them as input for construction of phylomemetic tree using UPGMA method as explained previously.

### 4.3.2 The Shortest Tree (*Maximum Parsimony*)

In constructing tree using this method, we can use Hamming distance as its basic construction process. This distance will show how much changes occurred upon two character sequences. The Hamming Distance between two sequence $x$ and $y$ with the same length is the number of different state for each same position for the two sequences. We can write it as $H(x, y) = |\{i : x_i \neq y_i\}|$. Normalized Hamming distance is another word for *Manhattan* distance, which have mutation value = 1, as we have previously explained.

For the cellphone memeplex we are about to infer, the Hamming distance between two memeplexes is the number of different allomeme between the two sequences. We can normalize it by dividing the value with the number of memes constituting its memeplex. The shortest distance of a tree is defined as minimum number of Hamming distance from the sequences constituting the tree.

By using data of Hamming distance between sequences, we try to construct a tree with the minimum Hamming distance. In this paper, we use *minimum spanning tree (MST)* technique by using Kruskal algorithm that approximately will give the minimum

Hamming distance (Situngkir, 2004b). MST is an algorithm to find the shortest path to connect one object with other objects in a system with certain number objects. Shortly MST with Kruskal Algorithm can be written as:

```
Begin
        do while ( all vertex in the graph)
                Find the least edge in the graph
                Mark it with any given color e.g. blue
                Find the least unmarked (uncolored) edge in the graph that
                doesn't close a colored or blue circuit
                Mark this edge red
        end;
The blue edges form the desire minimum spanning tree;

end
```

The expected output is a tree that presumably gives the tree with sum of all distance is smallest (minimum).

## 5. Result and Discussion

We have modeled cellphone as a system constituted of certain memes or memeplex that we determine from the features of the cellphones. From the data, we perform two simulations to construct the model of phylomemetic tree of cellphones. The first simulation is constructing phylomemetic tree using UPGMA method, while the second construct phylomemetic tree using the shortest tree.

In the first simulation, we tried to see how the evolutionary relationship formed by looking at the similarity of allomemes that constitute a cellphone. The result of this first simulation is resulting phylomemetic tree as seen in the figure 5 below. From the tree, we can gain the picture of which types of cellphones that has the close relationship with a certain cellphone. The closeness of the relationship designated by the phenomenon of clustering in the tree. Cellphones in the same cluster (designated by the same line color, e.g. red), will have close characteristic one with another and will share the same allomeme, while that cellphones in different clusters will have different characteristics and have different set of memeplex.

From the phylomemetic tree using UPGMA, there is an interesting result that is the way we estimate or approximate innovation or changes in which types of cellphones that will emerging new (or certain featured) cellphone (number and name of cellphones shown in appendix 2). This can be predicted by looking at which type of cellphone that lies in the same branching point. For example, if we look at the cluster with the red line, the 13$^{th}$-cellphone (series 7110), is approximately resulted from innovation of the 4$^{th}$-cellphone (series 6130) or the 5th (series 6150).

However, tree in Figure 5 cannot give sufficient information about the precise ancestor of a cellphone and when the ancestor diverge into new cellphone. This is understandable since the basic construction of the tree is merely based on its similar characteristics. But to say the least, the information of which cellphone related to it or not, can be seen as which cellphone evolves and which are not. From figure 5, we can see that the 1$^{st}$-cellphone and the 25$^{th}$, is innovatively less evolving compares to the others. It is because there is no other type related closely to them.

In the second simulation, we tried to construct cellphone phylomemetic using The Shortest Tree. The Hamming distance between cellphone memeplexes used as the inputs in this method. In this method we do not consider phylomemetic tree as a binary tree, so that is possible that one ancestor node may have one or more than one descendent.

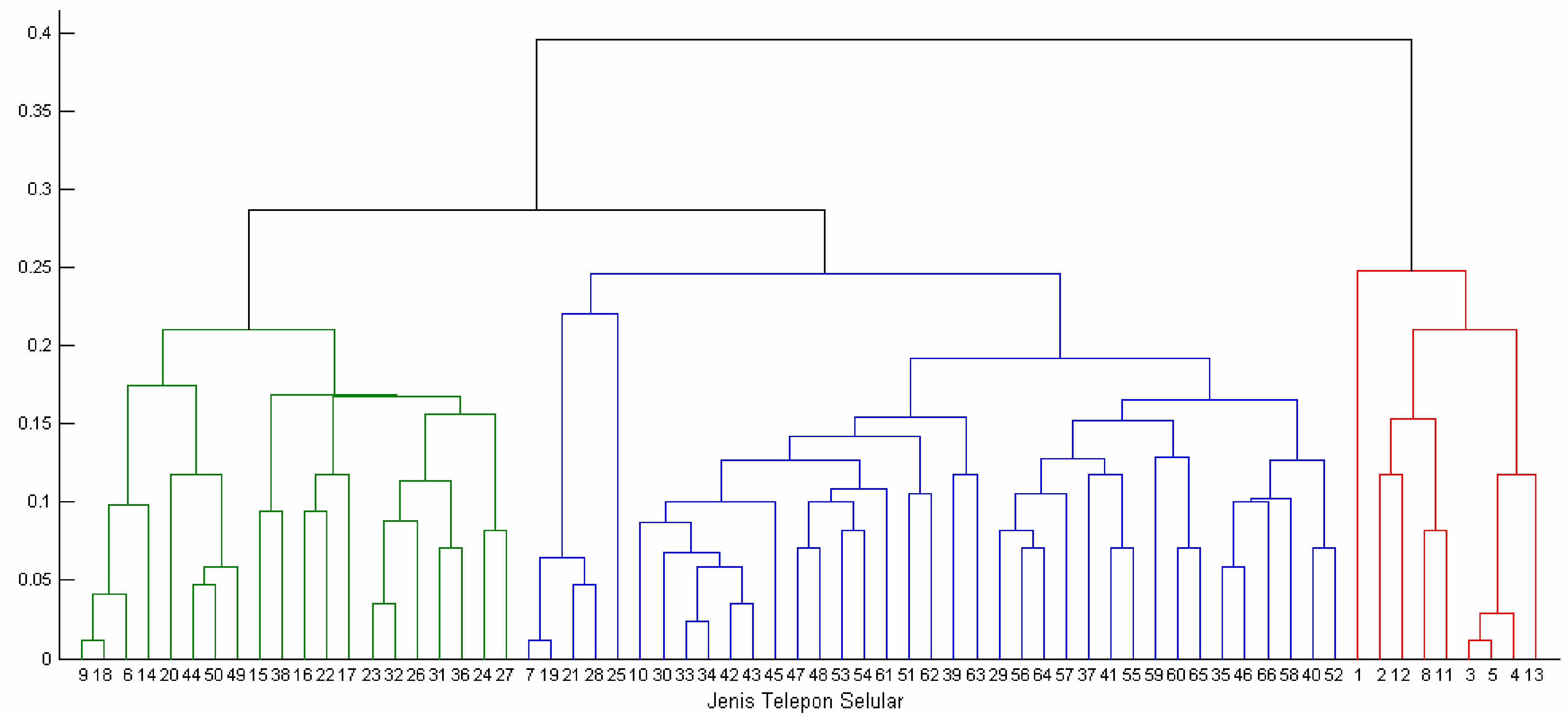

**Figure 5**
**Phylomemetic tree for Nokia using UPGMA method**

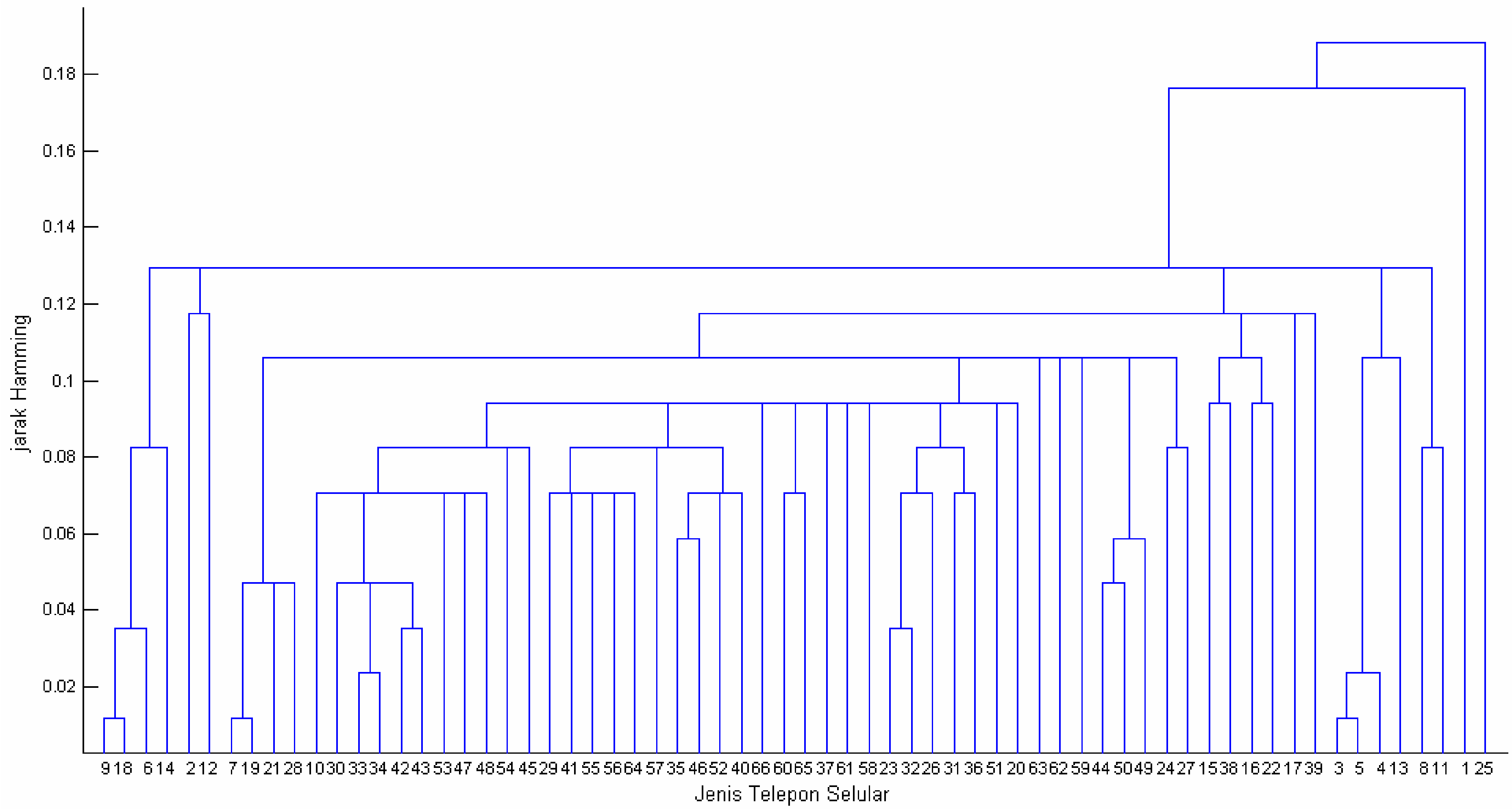

**Figure 6**
**Phylomemetic tree of Nokia using MST**

The result of the usage of this method is shown in Figure 6. By comparing the tree from UPGMA, we can see the quite different topology or branching pattern. This is of no wonder, since in UPGMA, we only take the smallest distance among species into account (local), while that in The Shortest Tree, it is a must to find the tree with the total distance is the smallest (locally and globally). Although so, in both trees, there are several similar clustering.

The tree generated from The Shortest Tree method gives a different picture on which type is the ancestor of a cellphone, rather than with the first tree. In this tree (Figure 6), it is approximated that the 25$^{th}$ and the 1$^{st}$ cellphone are the predecessor of other cellphones. Other cellphones emerge as the result of innovation and mutation from the cellphone memes. As well as in the first tree, in the second tree we can see that the close cellphone (appear in the same branching) will have similarity in terms of allomeme set, so that we can predict the appearance of cellphone as the innovation result from the relatively close cellphones.

From the Hamming distance among types (figure 7), we can see how close one type of cellphone with other types, where the bluer shows the similarity or higher closeness among cellphones. From the picture of Hamming distance among those cellphones, we can see that the 30$^{th}$ to the 66$^{th}$ have a close distance with each other, so that we can say there is a share of similar characteristics among them. Besides, the Hamming distance is showing an assumption that evolution of cellular telephone occurs by the mean of innovation, which is gradual and accumulative.

Other thing to see from Figure 7 is the presence of isolated cellphones, which means they do not have similarity or share the same characteristics. This can be seen from the wide Hamming distance (red) with other cellphone, e.g. the 1$^{st}$, the 8$^{th}$ and the 11$^{th}$ cellphone. It is assumed that those types of cellphones no longer emerging innovation, unlike other types. This is due to the fitness of types of cellphones with their environment.

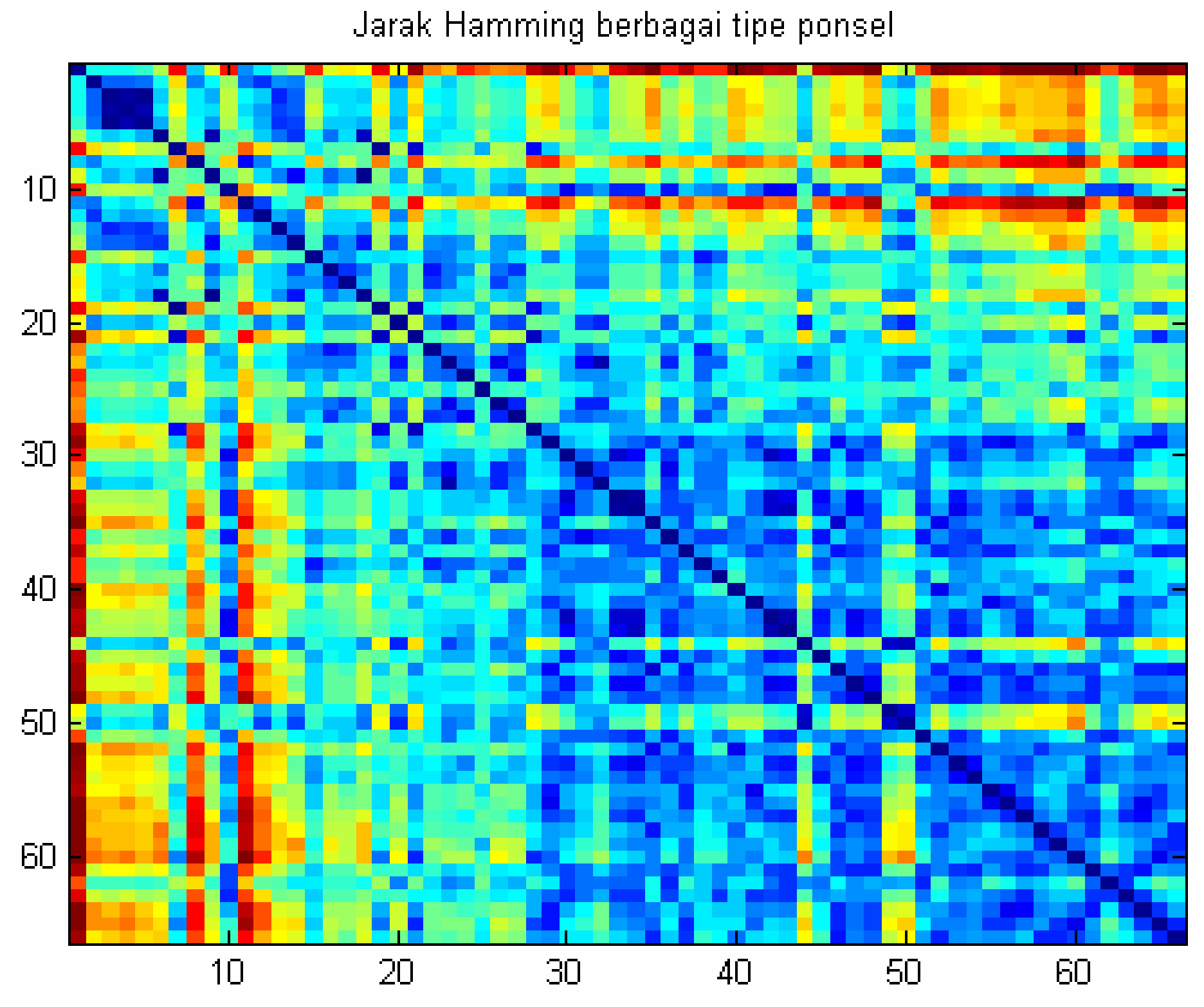

**Figure 7**

**Hamming distance from various cellphones**. The bluest area means the smallest distance (higher similarity), while red means the broader distance (the more different)

Beside constructing tree using the above methods, we also tried to visualize tree that showing the ancestral relationship by merely based on serial numbers and its time appearance[3], without looking at the characters they have. It is believed that a cellphone with certain serial number is resulted from the innovation of the previous cellphones with the same serial number. The resulting tree is shown in appendix 3. Generally, the branching pattern in this phylomemetic with the simulation-generated tree is relatively different. This shows innovation in cellphones either in design or technology, is not directly related with the cellphone serial number, so that it is very likely that a different serial number cellphones might have the same technology capacity.

The phylomemetic model we construct here, we should admit the need of further development to gain more thorough understanding about the process of innovation in technological artifacts. In phylomemetic analysis we perform here, we regard each allomeme to have the same contribution in determining the whole character of a technological artifact. In practical, the contribution of each allomeme in affecting the characteristics of technological artifact will be different. There might be certain allomemes, which significantly affect the whole character but there are also with small influence. In other words, there lies different contribution of *fitness value* – which is generally used to measure how far the adaptivity of an evolutionary system to their environment – over the fitness of the technological artifact from the constituting memes. Besides, one meme will not stand-alone. There is an epistetic factor, where one meme might be influenced by other memes. Advanced study and enriching data supply is a must to observe how far certain memes affect the fitness of a type of cellphone.

Other thing to concern is that the construction of phylomemetic tree we perform here is still a simple model in the way of analyzing innovation in technological artifacts. Of course we need further advanced model to show how trajectory that is through by an artifact from simple form artifact to the more complex in innovation process, and the dynamics of the emergence and the extinction of the technological artifacts along the evolution process.

To say the least that in this paper, we have showed you the possibility using evolutionary theory, in this case memetics, in analyzing innovation phenomena in technological design.

## 6. Concluding Remarks

Innovation is an evolutionary process. In effort of understanding innovation process in technological artifacts, we tried to view it from perspective of memetics. In this paper, we tried to show how phylomemetic analysis might become one of promising alternatives in understanding innovation of technological artifacts.

From the phylomemetic tree we worked in simulation, we can gain several interesting things in accordance with cellphone innovation. We gain a better picture of evolutionary relationship of cellphone based on allomeme similarity. Comparison of several methods shows different ancestral relationship. This is logical, since there lies different assumption used in constructing the method. Phylomemetic tree is showing memetic continuity relationship from one type of cellphone to others. Using Phylomemetic tree, we can show how and when a technological artifact diverges as the result of innovation, and which artifact that stop evolving.

Other thing to conclude is the way memetic works as an analytical tool to analyse technological innovation. Of course, phylomemetic is only the beginning step, which in turns should be followed up with other researches, in order to explain and understand the process in innovation of technological artifacts holistically. Other tools like agent-based model is necessary to build a more complete picture of this phylomemetic analysis.

[3] Data about serial number and time appearance can be obtained in the press release issued by Nokia in website http://www.nokia.com/nokia/0,8764,113,00.html

**Acknowledgements**

The authors would like to give sincere thanks to the researchers in BFI, for their support and discussions, and especially to Saras for her literary assistance. The authors also thank Surya Research Int'l. for financial support along the working hours of this paper.

## Appendix 1
## List of Constituting Allomeme of Cellphone Memeplexes

| No | Allomeme |
|---|---|
| 1 | Position of antenna |
| 2 | The presence of Handsfree |
| 3 | Headset |
| 4 | Screen Color |
| 5 | Screen size |
| 6 | Extra screen |
| 7 | Volume or the dimension of cellphone |
| 8 | Weight of cellphone |
| 9 | Cover modification |
| 10 | Clamshell type or not |
| 11 | The shape of cellphone |
| 12 | Number of parts |
| 13 | Position of cellular screen |
| 14 | Shape of Keypad/Keyboard |
| 15 | The presence of Keyboard |
| 16 | Illuminated Keypad |
| 17 | Joystick in keypad |
| 18 | Type of battery |
| 19 | Battery recharging Time |
| 20 | Call waiting |
| 21 | Call divert |
| 22 | Call timer |
| 23 | Call hold |
| 24 | Voicemail |
| 25 | Conference Call |
| 26 | Voice dialing |
| 27 | Appearing Name and caller number |
| 28 | SMS facility |
| 29 | Number of characters for SMS |
| 30 | Delivery Report for SMS |
| 31 | Prediction of input letter in SMS |
| 32 | Can transmit data |
| 33 | MMS facility |
| 34 | Facility to open and send E-mail |
| 35 | Browser Facility |
| 36 | Download Facility |
| 37 | Chatting Facility |
| 38 | Game facility |

| No | Allomeme |
|---|---|
| 39 | Modifiability in games |
| 40 | Calculator facility |
| 41 | Clock |
| 42 | Alarm |
| 43 | reminder note facility |
| 44 | Adress book facility |
| 45 | notepad |
| 46 | Calendar |
| 47 | stopwatch |
| 48 | Screensaver |
| 49 | Digital Camera |
| 50 | Radio AM/FM |
| 51 | Voice recorder |
| 52 | Video Recorder |
| 53 | Ability to receive pictures |
| 54 | High resolution picture |
| 55 | Edit Picture facility |
| 56 | Ringtone type |
| 57 | Vibration alert facility |
| 58 | Silent Mode |
| 59 | Keypad tone |
| 60 | Download ringtone facility |
| 61 | ringtone composer |
| 62 | Personalized ringtone |
| 63 | SMS Alert |
| 64 | Media Player |
| 65 | MS Office facility |
| 66 | Having infrared |
| 67 | Bluetooth |
| 68 | Data cable facility |
| 69 | Facility to keep Missed Call |
| 70 | Memory capacity |
| 71 | the presence of addable slot for memory |
| 72 | Time period (Live span) for active cellphone |
| 73 | Talk time |
| 74 | Water resistant |
| 75 | Shock Resistant |

| No | Allomeme |
|---|---|
| 76 | Dual/triple band facility |
| 77 | PIN number facility |
| 78 | Automatic Lock Keypad facility |
| 79 | Multilingual |
| 80 | GPRS (General Package Radio Service) |
| 81 | WAV Browser |
| 82 | HSCSC ((High-Speed Circuit-Switched Data) |
| 83 | Script **Java** |
| 84 | Converter for temperature, time, currency, etc. |

## Appendix 2
## Number and Cellphone Name Used in Phylomemetic Tree

| No | Type |
|---|---|
| 1 | Nokia RinGo™ |
| 2 | Nokia 5110™ |
| 3 | Nokia 6110™ |
| 4 | Nokia 6130™ |
| 5 | Nokia 6150™ |
| 6 | Nokia 8810™ |
| 7 | Nokia 9110 Communicator™ |
| 8 | Nokia 650™ |
| 9 | Nokia 8850™ |
| 10 | Nokia 6100™ |
| 11 | Nokia 640™ |
| 12 | Nokia 3210™ |
| 13 | Nokia 7110™ |
| 14 | Nokia 8210™ |
| 15 | Nokia 8910™ |
| 16 | Nokia 6210™ |
| 17 | Nokia 6250™ |
| 18 | Nokia 8890™ |
| 19 | Nokia 9110i Communicator™ |
| 20 | Nokia 3310™ |
| 21 | Nokia 9210 Communicator™ |
| 22 | Nokia 6310™ |
| 23 | Nokia 3330™ |
| 24 | Nokia 8310™ |
| 25 | Nokia 5510™ |
| 26 | Nokia 5210™ |
| 27 | Nokia 6510™ |
| 28 | Nokia 9210i Communicator™ |
| 29 | Nokia 7650™ |
| 30 | Nokia 7210™ |
| 31 | Nokia 3510™ |
| 32 | Nokia 3410™ |
| 33 | Nokia 6610™ |
| 34 | Nokia 6610i™ |
| 35 | Nokia 3650™ |
| 36 | Nokia 3510i™ |
| 37 | Nokia 6650™ |
| 38 | Nokia 6310i™ |
| 39 | Nokia 8910i™ |
| 40 | Nokia 3300 |
| 41 | Nokia 6800™ |
| 42 | Nokia 7250™ |
| 43 | Nokia 7250i™ |
| 44 | Nokia 2100™ |
| 45 | Nokia 5100™ |
| 46 | Nokia 6600™ |
| 47 | Nokia 6220™ |
| 48 | Nokia 6230™ |
| 49 | Nokia 2300™ |
| 50 | Nokia 1100™ |
| 51 | Nokia 3100™ |
| 52 | Nokia 7600™ |
| 53 | Nokia 3200™ |
| 54 | Nokia 5140™ |
| 55 | Nokia 6810 Messaging Device™ |
| 56 | Nokia 6820 Messaging Device™ |
| 57 | Nokia 7200™ |
| 58 | Nokia 7610™ |
| 59 | Nokia 7700 Media Device™ |
| 60 | Nokia 9500 Communicator™ |
| 61 | Nokia 3220™ |
| 62 | Nokia 2600™ |
| 63 | Nokia 2650™ |
| 64 | Nokia 6170™ |
| 65 | Nokia 6260™ |
| 66 | Nokia 6630 Phone ™ |

**Appendix 3**
**The Constructed Tree Based on Serial number and Appearance Time**

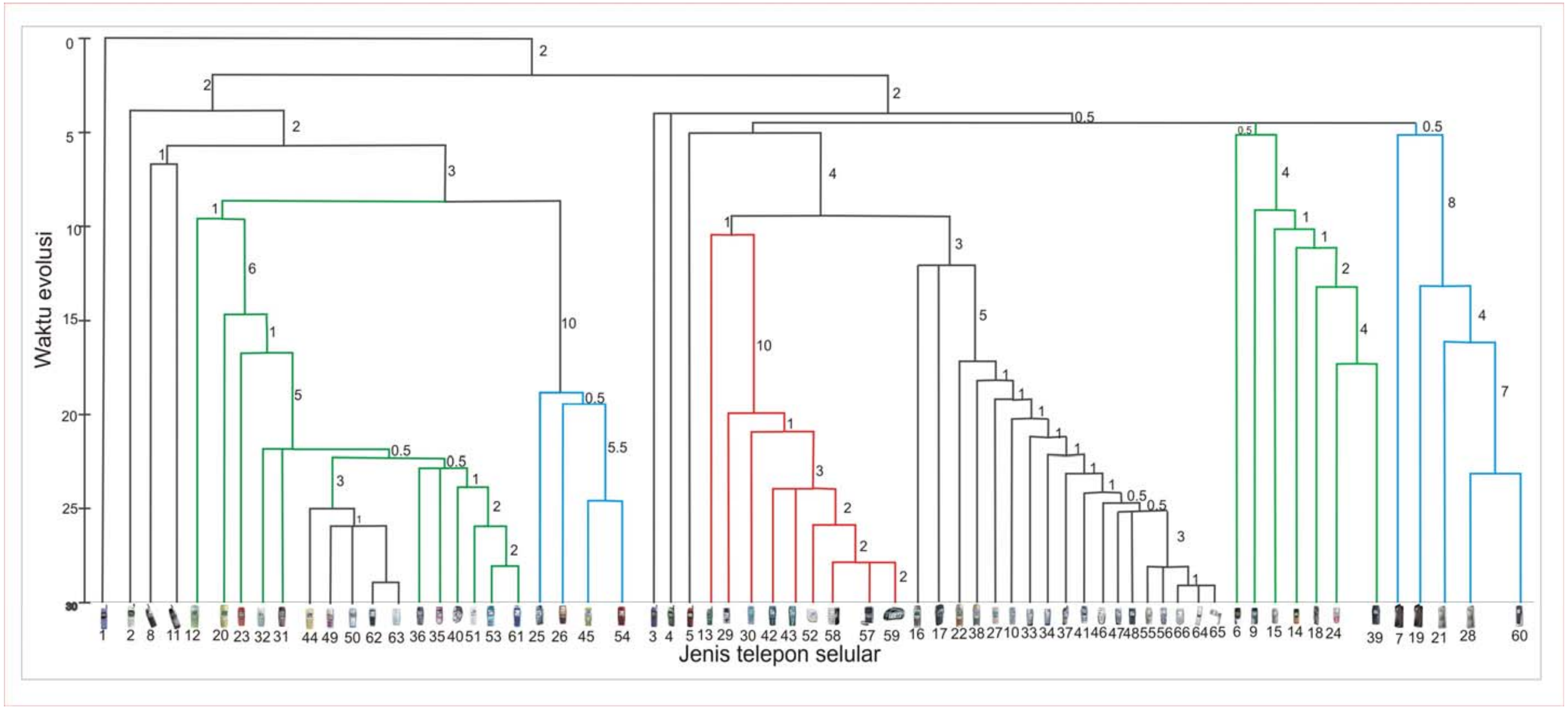